\documentstyle[a4,11pt]{article}
\begin{document}

\setlength{\baselineskip}{26pt}

\noindent
{\large \bf On partition function in Astronomy \& Astrophysics}

\vspace{4mm}

\setlength{\baselineskip}{18pt}

\noindent
{\Large M.K. Sharma$^1$, Monika Sharma$^1$ and  Suresh Chandra$^{2,3}$}

\noindent
$^1$School of Studies in Physics, Jiwaji University, Gwalior 474 011 (M.P.),
India

\noindent
$^2$Physics Department, Lovely Professional University, Phagwara 144411
(Punjab), India

\noindent
$^3$Zentrum f\"ur Astronomie und Astrophysik, Technische Universit\"at
Berlin, Hardenbergstrasse 36, D-10623 Berlin, Germany

\noindent
Emails: mohitkumarsharma32@yahoo.in; monika3273@yahoo.in;
suresh492000@yahoo.co.in

\vspace{6mm}

\noindent
{\bf Abstract.}
In order to analyze spectrum from the interstellar medium (ISM), spectrum of the 
molecule of interest is recorded in a laboratory, and accurate rotational and 
centrifugal distortion constants are derived. By using these constants, one can 
calculate accurate partition function. However, in the  same paper, where these constants are 
derived, the partition function is calculated by using a semi-empirical expression. 
 We have looked into the details of this semi-empirical expression and compared the 
values, obtained from it, with the accurate ones. As an example, we have considered 
the case of Methanimine (CH$_2$NH) which is detected in a number of cosmic objects. It
 is found that for the kinetic temperature $T > 120$ K, the semi-empirical expression 
gives large value as compared to the accurate one. The deviation becomes about 25\% larger than the accurate one at the 
kinetic temperature of 400 K. 
 
\vspace{2mm}

\noindent
{\bf Key words.} ISM: molecules; Partition function

\section{Introduction}

Partition function derived from the spectroscopic parameters connects to the
thermodynamic quantities. From time to time, in various contexts in Astronomy \& 
Astrophysics, computation of partition function has been a 
point of extensive discussion. For example, the calculation of partition function of 
H$_3^+$ has been in hot discussion in the context of solar continuum observations.
Several scientists calculated the partition function of H$_3^+$. The 
partition function of H$_3^+$ as a function kinetic temperature, calculated by 
Chandra {\it et al.} (1991), was declared by Sidhu {\it et al.} (1992) to be the best one.
Later on, Neale \& Tennyson (1995) extended the calculation to high temperatures.
The rotational partition function $Q$ is expressed as (Chandra {\it et al.}, 1991;
Sidhu {\it et al.}, 1992; Neale \& Tennyson, 1995)
\begin{eqnarray}
Q = \sum_i (2 J_i + 1) \ \mbox{exp}(-hc E_i/kT) \label{eq01}
\end{eqnarray}

\noindent
Here, $T$ is the kinetic temperature and the nuclear degeneracy is not considered. The
 $h$, $k$, $c$ are, respectively, the Planck constant, Boltzmann constant and speed of
 light. For a rotational level $i$ of asymmetric top molecule, the parameters 
$J_i$, $E_i$ are the rotational quantum number and energy, respectively. The 
$(2 J_i + 1)$ is the rotational statistical weight of level $i$. The values of $J_i$ 
and $E_i$ for a level can be calculated by using the rotational 
and centrifugal distortion constants for the molecule. 

However, in many papers where rotational and centrifugal distortion 
constants are derived with big efforts, 
the partition function is calculated by using the semi-empirical expression  
\begin{eqnarray}
Q = \sqrt{\frac{\pi}{ABC}\Big(\frac{kT}{h}\Big)^3} \label{eq02}
\end{eqnarray}

\noindent
where $A$, $B$ and $C$ are rotational constants. Details of this expression 
(\ref{eq02}) are discussed in the next section. In the present paper, we
 have discussed about the deviation obtained when one uses equation (\ref{eq02}) in 
place of (\ref{eq01}).  

\section{About equation (\ref{eq02})}

Kassel (1933) used partition function for calculation of thermodynamic quantities. 
Asymptotic expansion of the expression for the rotational partition function of a rigid 
symmetric top molecule for high temperature $T$ was given by Viney (1933) as the 
following.
\begin{eqnarray}
Q = \mbox{e}^{B h/4kT} \sqrt{\frac{\pi}{B^2A}\Big(\frac{kT}{h}\Big)^3}
\Big[1 + \frac{1}{12} \Big(1 - \frac{B}{A}\Big) \frac{B h}{kT} 
+ \frac{7}{480} \Big(1 - \frac{B}{A}\Big)^2 \Big(\frac{B h}{kT}\Big)^2 +
\ldots \Big] \label{eq03}
\end{eqnarray}

\noindent
where $A$ and $B$ are rotational constants for the symmetric top molecule. 

Since there is no explicit formula for energy of rotational level in case of an 
asymmetric top molecule, it is impossible to derive a rigorous asymptotic expansion 
for the partition function in this case. However, Gordon (1934) suggested that for a 
rigid asymmetric top molecule, in equation (\ref{eq03}), $B$ should be replaced by 
$\sqrt{BC}$ to get the expression as 
\begin{eqnarray}
Q = \mbox{e}^{\sqrt{BC}h/4kT} \sqrt{\frac{\pi}{ABC}\Big(\frac{kT}
{h}\Big)^3} \Big[1 + \frac{1}{12} \Big(1 - \frac{\sqrt{BC}}{A}\Big) 
\frac{\sqrt{BC} h}{kT} \hspace{10mm} \nonumber
\end{eqnarray}
\begin{eqnarray}
+ \frac{7}{480} \Big(1 - \frac{\sqrt{BC}}{A}\Big)^2 \Big(\frac{\sqrt{BC} h}
{kT}\Big)^2 + \ldots \Big] \hspace{-10mm} \label{eq04}
\end{eqnarray}

\noindent
Here, $A$, $B$ and $C$ are rotational constants for the asymmetric top molecule. This 
equation (\ref{eq04}) may be considered as a semi-empirical expression for a rigid
asymmetric top molecule. 

For sufficiently high temperature (or small rotational
 constants), equation (\ref{eq04}) reduces to  (\ref{eq02}), as the argument
 of the exponential would be very small, and the contributions of second, third, etc.
terms would be neglected. This expression (\ref{eq02}) is thus a semi-empirical 
relation derived for a rigid asymmetric top molecule and is valid at sufficiently high
 temperature. Moreover, the expression (\ref{eq04}) is 
better than  (\ref{eq02}). But, this expression (\ref{eq02}) is being 
used regularly by the scientists. Probably they are 
not aware of the facts about the expression (\ref{eq02}), and as a tradition, this
expression is being used. 

\section{Results and discussion}

Now, we have three equations (\ref{eq01}), (\ref{eq02}) and (\ref{eq04}), for the
partition  function.  Let us denote their values as $Q_{1}$, $Q_{2}$, $Q_{3}$, 
respectively, for convenience. There are many papers where accurate rotational and 
centrifugal distortion constants of asymmetric 
top molecule are derived, and in the same paper the partition function is calculated 
with the help of equation (\ref{eq02}). Some examples of such papers are: Widicus {\it et al.} (2003) (for NH$_2$CH$_2$CH$_2$OH), Widicus Weaver {\it et al.} (2005) (for 
CHOCH$_2$OH), Motoki {\it et al.} (2014) (for CH$_2$NH). We have not intention about 
them. We have considered the case of Methanimine (CH$_2$NH) and have used the data of Motoki {\it et al.} (2014), given 
in Table 1.

Using the values given in Table 1, we have calculated accurate energies of rotational 
levels with the help of the computer code ASROT (Kisiel, 2001). For calculation of
partition function, we have considered the levels having energy up to 1475 cm$^{-1}$, 
which is sufficiently large  for the kinetic temperature up to 400 K, considered here.
 Using these energies of levels, the partition function is calculated with the help of
 equation (\ref{eq01}). At 400 K, the contribution of the highest level considered is 
less than 0.01\%. It shows that the considered set of energy levels is sufficient, and further high 
energy levels would not contribute to the partition function significantly, up to a 
temperature of 400 K.

The results obtained $Q_{1}$ and $Q_{2}$ are presented in Figure \ref{fig1} as a 
function of kinetic temperature $T$ in the gas. The values of partition function 
reported by Motoki {\it et al.} (2014) were obtained after multiplying by a factor of 
3 for the nuclear degeneracy. Therefore, they differ by a factor of 3 as compared to 
those obtained with the help of equation (\ref{eq02}).

As obvious, the $Q_{3}$ is always larger than $Q_{2}$. For the 
temperature $T < 120$ K, the deviation of $Q_2$ with respect to $Q_{1}$ is less than 
1\%, and $Q_{2}$ is always smaller than $Q_{1}$. For $T > 120$ K, the deviation of 
$Q_{2}$ with respect to $Q_{1}$ increases with temperature. The deviation 
becomes 25\% at $T = 400$ K. It is quite large, and the values of thermodynamic 
quantities would change significantly. 

The contribution of centrifugal distortion constants to partition function increases when we go to the
 levels with large value of $J$. In some molecules, the 
contribution even may change the sequence of levels (Sharma 
{\it et al.,} 2014). It is interesting to note that equation (\ref{eq02}) does not
give accurate results even at high temperatures.

\section{Conclusion}

It may finally be concluded that when the  rotational and centrifugal distortion 
constants are available, the partition function should be calculated with the help of
equation (\ref{eq01}). It may require some efforts, but the reliable results would be 
available in literature.

\vspace{5mm}

\noindent
{\bf Acknowledgments}  

Suresh Chandra thankfully acknowledges nice hospitality of Prof. Dr. D. 
Breitschwerdt and Prof. Dr. W.H. Kegel of Technical University, Berlin, Germany,
 where this investigation was finalized. He is grateful to the Alexander von 
Humboldt Foundation, Bonn (Germany), for financial support and to the Lovely 
Professional University, Punjab (India) for support and encouragement.

\vspace{6mm}

\noindent
{\large\bf  References}

\begin{description}

\item{} Chandra,  S., Gaur, V.P. and  Pande, M.C: 1991, J. Quant. Spectrosc. Rad. 
Trans. {\bf 45}, 57 

\item{} Gordon, A.R.: 1934, J. Chem. Phys. {\bf 2}, 65.

\item{} Kassel, L.S.: 1933, J. Chem. Phys. {\bf 1}, 576.

\item{}  Z. Kisiel, in: J. Demaison,  {\it et al.}
(Eds.), {\it Spectroscopy from Space}, Kluwer, Dordrecht, 2001, pp. 91.

\item{} Motoki, Y., Isobe, F., Ozeki, H. and Kobayashi, K.: 2014, Astron. Astrophys. 
{\bf 566}, A28. 

\item{} Neale,  L. and  Tennyson, J.: 1995, Astrophys. J. {\bf 454}, L169.

\item{} Sharma, M., Sharma, M.K.,Verma,  U.P. and Chandra, S.: 2014, Adv. Space Res. 
{\bf 54}, 252.

\item{} Sidhu,  K.S., Miller, S. and Tennyson, J.: 1992, Astron. Astrophys. {\bf 255},
 453. 

\item{} Viney, I.E.: 1933, Proc. Cambr. Phil. Soc. {\bf 29}, 142. 

\item{} Widicus, S.L., Drouin, B.J., Dyl, K.A. and Blake, G.A.: J. Mol. Spectr. 
{\bf 217}, 278.

\item{} Widicus Weaver, S.L., Butler, R.A.H., Drouin, B.J., Petkie, D.T.,
Dyl, K.A. and De Lucia, F.C.: 2005, Astrophys. J. Suppl. {\bf 158}, 188.

\end{description}

\newpage

\vspace{4mm}

\hspace{-7mm}
\begin{tabular}{lr|lr}
\multicolumn{4}{l}{Table 1. Rotational and centrifugal distortion constants in MHz.} \\
\hline
\multicolumn{1}{c}{Parameter} & \multicolumn{1}{c|}{Value} & \multicolumn{1}{c}
{Parameter} & \multicolumn{1}{c}{Value} \\
\hline
$A$ & $196210.87629$ & $H_{JK}$ & $2.6068 \times 10^{-6}$ \\
$B$ & $34641.699423$ & $H_{KJ}$ & $6.125 \times 10^{-6}$  \\
$C$ & $29351.496787$ & $H_{K}$ & $877.75 \times 10^{-6}$ \\
$D_{J}$ & $56.00411 \times 10^{-3}$ & $h_{1}$ & $33.067 \times 10^{-9}$ \\
$D_{JK}$ & $598.6775 \times 10^{-3}$ & $h_{2}$ & $32.462 \times 10^{-9}$ \\
$D_{K}$ & 6.383483 & $h_{3}$ & $9.091 \times 10^{-9}$ \\
$d_{1}$ & $-9.973179 \times 10^{-3}$ & $L_{K}$ & $-163.0 \times 10^{-9}$ \\
$d_{2}$ & $-2.007905 \times 10^{-3}$ & $l_1$ & $-0.541 \times 10^{-12}$ \\
$H_{J}$ & $0.01736 \times 10^{-6}$ & $l_3$ & $-0.472 \times 10^{-12}$ \\
\hline
\end{tabular}

\begin{figure}[h]
\vspace{0mm}
\vspace{10mm}
\caption{\small Variation of partition function $Q$ versus kinetic temperature $T$. 
Solid line is for $Q_1$, obtained from equation (\ref{eq01}) and the dotted line for 
$Q_2$, obtained from equation (\ref{eq02}).  }
\label{fig1}
\end{figure}

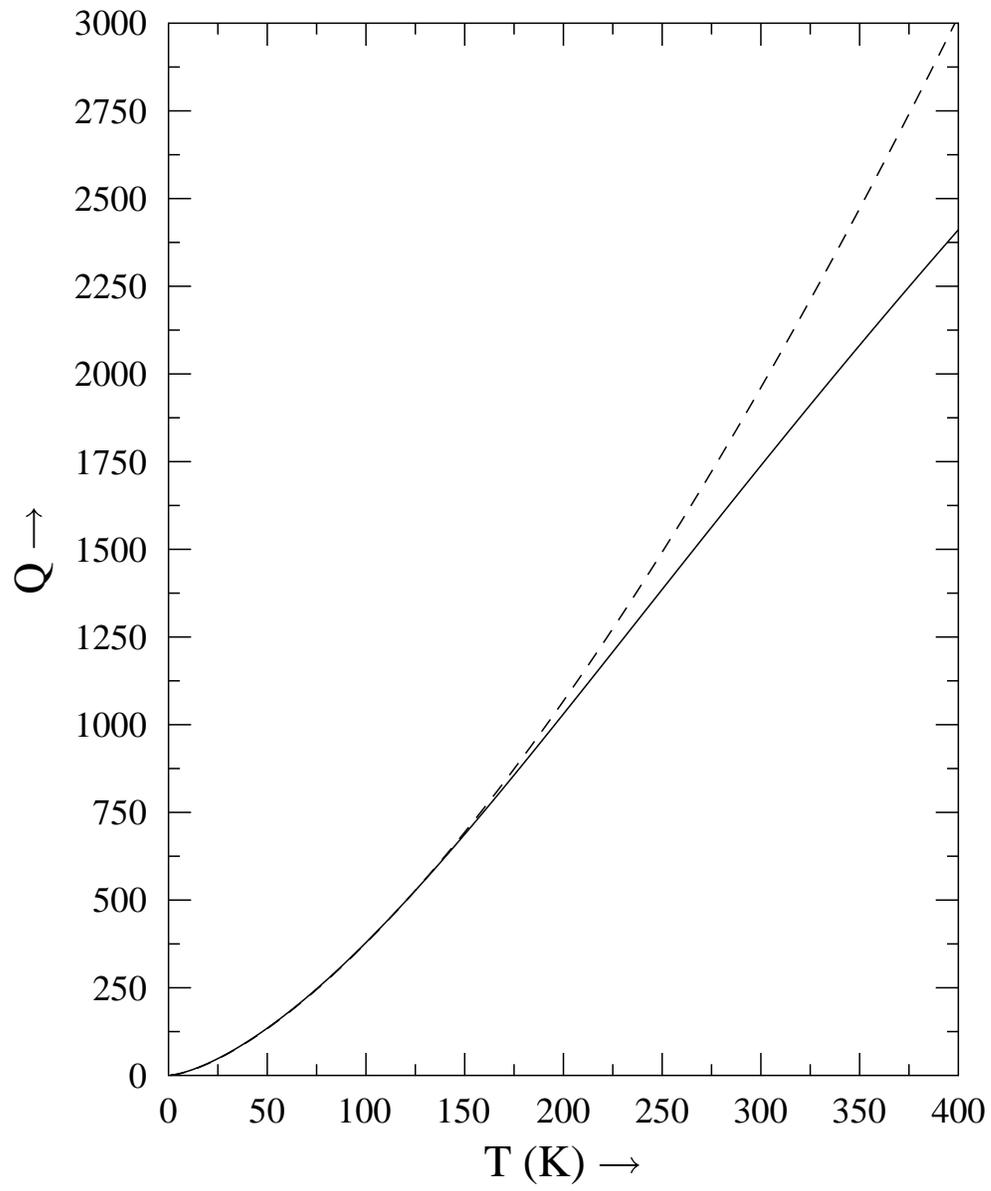
\end{document}